
\font\re=cmr8
\font\rn=cmr9

\def\narr{\advance\leftskip by 1.5 pc \advance\rightskip by 1.5 pc}

\hsize=6.1 true in
\vsize=8.7 true in
\nopagenumbers

\topinsert
\vskip 0.3 true in
\endinsert

\centerline{\bf BLACK HOLES, THE WHEELER-DE WITT EQUATION}
\centerline{\bf
AND THE SEMI-CLASSICAL APPROXIMATION}
\vskip0.4 true in
\centerline{MIGUEL E. ORTIZ}
\vskip 0.02 true in
\centerline{\re Institute of Cosmology,
Department of Physics and Astronomy,}
\centerline{\re
Tufts University, Medford, MA 02155, USA.}
\vskip 0.35 true in
\midinsert
\centerline{ABSTRACT}
\vskip 2pt
\re\baselineskip=10pt
\narr
The definition of matter
states on spacelike hypersurfaces of a 1+1 dimensional black
hole spacetime is considered. Because of small quantum
fluctuations in the mass of the black hole,
the usual approximation of
treating the gravitational field as a classical background
on which matter is quantized, breaks down near the black
hole horizon. On any hypersurface that
captures both infalling matter near the horizon and Hawking
radiation, a semiclassical calculation is
inconsistent. An estimate of the size of correlations
between the matter and gravity states shows that they are so
strong that a fluctuation in the black hole mass of order
$e^{-M/M_{Planck}}$ produces a macroscopic change in the
matter state.
(Based on a talk given at the 7th Marcel Grossmann Meeting on
work in collaboration with E. Keski-Vakkuri, G. Lifschytz and S. Mathur.)
\endinsert
\vskip 0.15 true in
\baselineskip 13pt

Various authors have recently argued that the information paradox
in black hole evaporation could be resolved if there is a mechanism
that allows the
information falling into a black hole to be transferred to the
outgoing radiation at the black hole horizon. With the additional
assumption that an infalling observer sees nothing unusual occuring
at the horizon, it follows that observations of the Hawking radiation
at future null infinity and of infalling matter
at the black hole horizon must be
complementary in some sense$^1$.
This has been expressed by 't Hooft$^2$ as the
statement that operators corresponding to
each observation must be strongly non-commuting despite the fact that
the two regions of interest, the
horizon and future null infinity, are
spacelike separated in the usual background
geometry of an evaporating
black hole.
Calculations in 1+1 dimensions seem to support this point of
view$^3$,
although assumptions are needed about a boundary condition
at the region of strong coupling.

The usual derivation of the black hole paradox assumes
a picture of quantum fields in a classical
background spacetime. This assumption can be justified by appealing to
a foliation of the black hole
spacetime which avoids all regions of strong curvature (or strong coupling)
and which includes late time hypersurfaces which capture the outgoing radiation
and the infalling matter close to the horizon. A hypersurface of
this last kind will be referred to as an S-surface (the potential
importance of these kinds of surfaces has been emphasized by
Susskind$^1$).

To ask whether a semi-classical description of black hole physics is
trustworthy, it is necessary to consider a quantum theory of gravity and
matter, and to understand how the semi-classical approximation is
obtained$^4$.
The Wheeler-DeWitt equation can be expanded
in the Planck mass, leading to a natural separation between a
quasi-classical gravitational state and a quantum matter state satisfying
the functional Schr\"odinger equation in the gravitational background. This
separation bewteen classical and quantum variables is thought to be reliable
whenever the inherenet scales in a problem are small compared
to the Planck scale.
For example, close to a black hole horizon, where part of an S-surface is
located, the energy density is relatively small, as are all co-ordinate
invariant local quantities. However, it was suggested
in Ref. 5 that the copious particle production associated with
the black hole evaporation process might lead to excessive decoherence of the
gravitational field, spoiling it's classical behaviour. It will be explained
below that a
careful analysis of the semi-classical approximation shows that
other scales in the black hole problem, not just the local invariants, can
spoil the expansion.  It turns out to fail precisely when one tries
to obtain the state of matter on an S-surface.
Below I shall briefly summarize the key results of Ref. 6, where a
detailed calculation can be found
that shows the breakdown of the semi-classical approximation
in the 1+1 dimensional CGHS model.

Working in 1+1 dimensions,
the configuration space of the gravitational field is the space of
1-geometries. A 1-geometry is an equivalence
class of metrics under the action of spatial diffeomorphisms.
In the model that we consider in
Ref. 6, where a dilaton is present as part of the gravitational
field, a 1-geometry is defined by the function $\phi(s)$ giving the dependence
of the dilaton field on the proper distance along a hypersurface. This must be
supplemented by a boundary condition giving the point of origin of the
proper distance.

If we include a matter field, we can express the configuration space of matter
fields as the space of functions $f(s)$. A solution to the
Wheeler-DeWitt equation including matter then takes the form
$$
\Psi[\phi(s), f(s)]
\eqno(1)
$$
We wish to approximate this functional by a solution
$$
\psi_{\cal M}[f(s),\phi(s)]
\eqno(2)
$$
of the functional Schr\"odinger equation on a mean background spacetime
${\cal M}$ times a quasi-classical gravitational state approximating ${\cal M}$
and depending only on $\phi(s)$. Note that given a spacetime ${\cal M}$, the
function $\phi(s)$ (plus an appropriate boundary condition) specifies a
unique timeslice in ${\cal M}$, and so acts like the time variable for the
Schr\"odinger picture state, just as $s$ acts like a spatial coordinate;
the boundary condition for an asymptotically
flat spacetime is just the location of the hypersurface at spacelike infinity
which itself defines the zero-point of the proper distance.

For the black hole, we make a separation between the matter forming the black
hole, which we include as part of the quasi-classical variables, and the
fields that give rise to the Hawking radiation which are denoted by the field
$f$. Regardless of how the black hole is formed, the quasi-classical state
representing the classical variables cannot be exactly classical.
For example, the infalling matter cannot be
in an exact energy eigenstate. If it is to be sufficiently localised to form a
black hole, there must be a spread in its energy of at least $1/M$ where
$M$ is the mass of the black hole in Planck units. Thus the background
spacetime
has a mass that is uncertain by at least $1/M$.
When we write the Schr\"odinger
equation, it is important to remember these Planck scale uncertainties in
the mean background
spacetime. If the semi-classical approximation is reliable, the
uncertainties should not affect matter propagation. There is a simple criterion
for determining whether or not this is the case. Consider two different choices
for the mean background spacetime which are solutions with masses that
differ on the order of $1/M$ or below. A priori each solution can be taken as a
mean background for the semi-classical approximation.
If we are to write the state of gravity and matter as the product of
a gravitational state, which contains both mean spacetimes in its spread,
and a single Schr\"odinger type matter state, then it must make no
difference to the matter state which of the two mean spacetimes we take to
be the background.

Let us make this statement more precise. We are interested in obtaining a
functional giving the state of matter on 1-geometries.
Consider the two mean spacetimes described above and
any 1-geometry that embeds in both.
Suppose that we are given a natural
way of comparing the matter states (defined using quantum field theory)
on each of the spacetimes, through
the common 1-geometry. Then the
criterion for semi-classicality is that the two matter states
be virtually indistinguishable when compared on
any choice of common 1-geometry.
Suppose we take as an intial condition that the two states be
indistinguishable at past null infinity. There is no guarantee
that this will continue to be true under
dynamical evolution of the states in the respective
spacetimes. In all familiar physical situations the states do remain
indistinguishable, meaning that quantum gravity effects can be ignored.
However, a black hole is different: by the time states are
evolved to the type of common 1-geometry which
we have called an S-surface, the states become radically different.
This result shows that the quantum state of a
black hole cannot be approximated
as a quasi-classical gravitational state times a single matter state
defined using qauntum field theory on the gravitational background.

In Ref. 6 details are given of how to indentify 1-geometries in
different spacetimes, and how to compare states through an inner product that
is intrinsic to the 1-geometry. The identification of 1-geometries involves the
boundary condition that at spatial infinity the states should always be
indistinguishable, which is a physical constraint corresponding to a preferred
role for a strictly classical observer at infinity.
The inner product is
constructed through a mode decomposition in terms of the proper distance along
the 1-geometry and is in this sense unique. The essential feature of an
S-surface that leads to the difference in the states is that there is a large
shift in the location of that surface in adjacent spacetimes.
In terms of Kruskal coordinates near the horizon
the shift is of order $e^{M}$ in Planck units, a huge number,
for a Planck scale difference in mass between the spacetimes. This shift is
macroscopic and is more than sufficient to ensure that
two states that began life at past null infinity with an overlap of
order unity,
$$
\left\langle \psi_{\cal M}\left[f(s)\vert\phi_{{\cal I}^-}
(s)\right]\;\vert\; \psi_{{\cal M}+\delta{\cal M}}
\left[f(s)\vert\phi_{{\cal I}^-}(s)\right]\right\rangle\sim 1
$$
have virtually zero overlap on the S-surface:
$$
\left\langle \psi_{\cal M}\left[f(s)\vert\phi_{\Sigma_S}
(s)\right]\;\vert \;\psi_{{\cal M}+\delta{\cal M}}
\left[f(s)\vert\phi_{\Sigma_S}(s)\right]\right\rangle\sim 0
$$
Indeed even for a difference in masses of order $e^{-M}$, the overlap on an
S-surface is close to zero. It is interesting to note that this
scale leads to an entanglement entropy between the geometry and
the matter on the S-surface of
the same order as the usual black hole entropy.

We can deduce from these results that fluctuations in geometry prevent one from
using quantum field theory on curved spacetime to determine the state of matter
on an S-surface. Consequently, there is no way to compute both the Hawking
radiation at future null infinity and the backreaction at the horizon. This
strongly suggests that we should consider the entire solution to the
Wheeler-DeWitt equation if we are to understand the physics of black holes,
and not just its semi-classical projection.

\vskip 0.4 true in
\centerline{\bf Acknowledgements}
\vskip 0.1 true in
This paper summarises the results of
a collaboration with Esko Keski-Vakkuri, Gilad
Lifschytz and Samir Mathur, and I would like to thank my collaborators
and the organisers
for the opportunity to communicate the work at MG7.
A more complete version of this work with a full list of acknowledgements
may be found in Ref. 6, and some early motivations behind
the work are described in Ref. 5.

\vskip 0.4 true in

\noindent{\bf References}
\vskip 0.1 true in
{\rn
\item{1.}  L. Susskind, L. Thorlacius and J. Uglum, {\sl
Phys. Rev. } {\bf D48} (1993) 3743;
L. Susskind, {\sl Phys. Rev.} {\bf D49}
(1994) 6606.

\item{2.} G. 't Hooft, {\sl Nucl. Phys.} {\bf B256}
(1985) 727;
G. 't Hooft, {\sl Nucl. Phys.} {\bf B335}
(1990) 138;
C.~R.~Stephens, G.~'t Hooft and B.~F.~Whiting,
{\sl Class. Qu. Grav.} {\bf 11} (1994) 621.

\item{3.} E.~Verlinde and H.~Verlinde, {\it A Unitary
S-matrix for 2D Black Hole Formation and Evaporation,}
Princeton Preprint, PUPT-1380, IASSNS-HEP-93/8,
hep-th/9302022 (1993), K. Schoutens, E.~Verlinde, and
H.~Verlinde, {\sl Phys. Rev. }{\bf D48} (1993) 2670.

\item{4.} See the review by C. Kiefer,
{\it The Semiclassical Approximation to
Quantum Gravity}, Freiburg University Report No. THEP-93/27,
to appear in {\it Canonical Gravity - from Classical to
Quantum}, edited by J.~Ehlers and H.~Friedrich (Springer,
Berlin 1994) (gr-qc/9312015).

\item{5.}  S.~D.~Mathur, {\it Black Hole Entropy and
the Semiclassical Approximation}, MIT report No. CTP-2304
(hep-th/9404135) (Invited Talk given at the International
Colloquium on Modern Quantum Field Theory II at TIFR
(Bombay), January 1994).

\item{6.} E. Keski-Vakkuri, G. Lifschytz, S. Mathur and M. E. Ortiz,
{\it Breakdown of the semi-classical approximation at the black hole
horizon}, hep-th/9408039.
}
\end